\documentclass[preprint,12pt]{elsarticle}

\usepackage{amssymb}
\usepackage{booktabs}
\usepackage{amsmath}
\usepackage{siunitx}
\usepackage{siunitx}  % for \num, \SI
\usepackage{booktabs} % for \toprule \midrule \bottomrule
\usepackage{float}    % for [H] placement
\usepackage{graphicx} % for \includegraphics
\usepackage{multirow} % for multirow in tables
\usepackage{booktabs}   % \toprule \midrule \bottomrule
\usepackage{pifont}      % 提供 \ding 命令
\newcommand{\cmark}{\ding{51}}   % ✓  \ding{51}
\newcommand{\cfdmark}{\ding{72}} % ★  \ding{72}
\usepackage{booktabs} 
\usepackage{enumitem}  % for [label=(\roman*)]
\usepackage{pifont}    % for \ding symbols

\usepackage{tikz}
\usepackage{subcaption}
\usepackage{tabularx}

\usepackage{pgfplots}
\pgfplotsset{compat=1.18}
\usepackage{pgfplotstable}
\usepackage{siunitx} % 可选，用于数值格式

\usetikzlibrary{arrows.meta, positioning}

\journal{Computer and Fluids}

\begin{document}

\begin{frontmatter}

%% Title, authors and addresses

%% use the tnoteref command within \title for footnotes;
%% use the tnotetext command for theassociated footnote;
%% use the fnref command within \author or \affiliation for footnotes;
%% use the fntext command for theassociated footnote;
%% use the corref command within \author for corresponding author footnotes;
%% use the cortext command for theassociated footnote;
%% use the ead command for the email address,
%% and the form \ead[url] for the home page:
%% \title{Title\tnoteref{label1}}
%% \tnotetext[label1]{}
%% \author{Name\corref{cor1}\fnref{label2}}
%% \ead{email address}
%% \ead[url]{home page}
%% \fntext[label2]{}
%% \cortext[cor1]{}
%% \affiliation{organization={},
%%             addressline={},
%%             city={},
%%             postcode={},
%%             state={},
%%             country={}}
%% \fntext[label3]{}

\title{A fast Physics-Informed Neural Networks based approach to the 2D design of turbine blades}

\author[rub]{Yuan Huang}
\ead{yuan.huang@ruhr-university-bochum.de}  

\author[rub]{Francesca Di Mare\corref{cor1}}
      
%% 2) 
\cortext[cor1]{Corresponding author.}

%% 3) 
\affiliation[rub]{%
  organization = {Ruhr University Bochum, Chair of Thermal Turbomachines and Aero Engines},%
  addressline  = {Building IC, Room E2-63, Universitaetsstr. 150},%
  city         = {Bochum},%
  postcode     = {44780},%
  country      = {Germany}%
}

%% Abstract
\begin{abstract}
Rapid aerodynamic screening of turbomachinery blades across wide operating envelopes remains a major computational bottleneck in preliminary design, particularly for energy-conversion and storage systems such as emerging Carnot batteries. Physics-informed neural networks (PINNs) offer a mesh-free alternative to conventional CFD, yet convergence and accuracy often deteriorate for complex blade geometries and off-design flows.  We propose a progressive Euler-PINN framework that (i) gradually relaxes boundary conditions from tunnel flow without a blade to full outlet static pressure, and (ii) employs a geometry-aware dynamic loss-weighting scheme that intensifies residual penalties near highly curved boundaries. To the best of our knowledge, this is the first study to deploy a single PINN workflow for large-scale, engineering-grade screening of turbomachinery blade families across multiple operating conditions, covering ten NACA6 variants and 30 subsonic operating points. The proposed framework achieves CFD-comparable accuracy for pressure and velocity fields while reducing the computational cost required for family-wide blade screening. These results establish the method as a practical surrogate for two-dimensional turbomachinery blade pre-design and optimisation.
\end{abstract}

%%Graphical abstract
\begin{graphicalabstract}
\end{graphicalabstract}

%%Research highlights
\begin{highlights}
\item Research highlight 1
\item Research highlight 2
\end{highlights}

%% Keywords
\begin{keyword}
%% keywords here, in the form: keyword \sep keyword

%% PACS codes here, in the form: \PACS code \sep code

%% MSC codes here, in the form: \MSC code \sep code
%% or \MSC[2008] code \sep code (2000 is the default)
Aerodynamics, Physics-Informed, Neural Networks, Turbine Machine
\end{keyword}

\end{frontmatter}

%% Add \usepackage{lineno} before \begin{document} and uncomment 
%% following line to enable line numbers
%% \linenumbers

%% main text
%%

\section{Introduction}

Turbomachinery pre-design often calls for the aerodynamic screening of \emph{tens to hundreds} of two-dimensional airfoil variants over wide Reynolds--incidence matrices. Even in low-Mach regimes ($M_\infty<0.8$), a grid-converged Euler or low-Re RANS analysis of a single NACA profile still needs $\mathcal{O}(10^{6})$ boundary-layer-resolved cells \cite{Swanson2016}, and, judging from NASA's \textit{PALMO} statistics, about 16.4 CPU-h on modern 64-core clusters \cite{PALMO24}. PALMO itself ran $52\,480$ such cases for only 16 NACA-4-series airfoils, consuming $\approx8.6\times10^{5}$ CPU-h in total — a workload far beyond typical industrial timelines. Classical surrogate models (e.g.\ Kriging or radial-basis networks\cite{jouhaud2007surrogate,liu2017quantification}) amortise this cost once trained, yet they demand \emph{thousands} of high-fidelity snapshots and still extrapolate poorly outside the training hull \cite{Li2022Survey}.  

The energy-conversion role of turbomachinery in thermal-energy systems such as the emerging Carnot batteries (CB) underscores the importance of rapid, reliable blade and flow screening under diverse operating points. Modern CB systems are inherently multi-parameter and involve turbomachinery that must operate robustly across a wide range of operating conditions; they must optimize not only geometry but also component configurations, working fluids, thermal-storage media, cycle conditions, and part-load operations. The foundational review by Liang et al. documents that CB systems rely on tightly coupled subsystems — compressors, expanders/turbines, thermal-energy storage (TES), heat exchangers and working fluid networks — and that the technical performance and cost of the whole system critically depend on the component characteristics and their interactions \cite{Liang2022}. Other overviews of CB technologies likewise highlight that different cycle types (e.g.\ Rankine, Brayton, compressed-air, latent- or sensible-heat TES, etc.) impose widely varying demands on component design, control strategy, and fluid/thermal routing \cite{Dumont2020}. Therefore, a robust, geometry- and condition-aware surrogate (e.g.\ our boundary-moving PINN framework) — capable of handling sector-wide design spaces and cycles operating across multiple operating conditions — may significantly accelerate techno-economic trade-off analysis and preliminary design of CB components and subsystems.

Physics-informed neural networks (PINNs) remove the meshing bottleneck by embedding the governing PDEs directly in the loss; early single-airfoil demonstrations already show sub-percent $C_p$ errors without CFD data \cite{Cao2024PINN, huang2025physics}. However, scaling such solvers from one or two canonical profiles to an \emph{industrial} screening loop of dozens of airfoils remains an unresolved challenge: each new geometry triggers fresh optimisation, training time balloons accordingly, and convergence often stalls near highly-cambered sections where residual gradients become ill-conditioned \cite{Krishnapriyan2021, Rathore2024}. Bridging this gap — namely, delivering a PINN framework capable of evaluating an entire family of sub-sonic blades within the wall-clock time of a single CFD run — forms the central motivation of this study.

Physics-informed neural networks (PINNs) solve PDEs by minimising domain residuals and boundary errors within a neural-network loss \cite{Raissi2019}, and recent single-airfoil studies already reach sub-percent \(C_p\) errors in sub-sonic Euler flow with \(\mathcal{O}(10)\)-min GPU training \cite{Cao2024PINN}. However, systematic studies on complex or high-Re / turbulent flows reveal significant challenges: for instance, even modified PINN frameworks for turbulent or multiphysics flow report nontrivial prediction errors and require careful tuning and model/closure-choice \cite{Cai2021PINNFluid, Chuang2022ExperiencePINN}. More comprehensive reviews of PINN applications in fluid dynamics — covering turbulence, multiphase flows, multiphysics coupling, complex geometries and multi-regime flows — show that robustness, generalisation over varied geometries and flow regimes, and stable convergence remain unsolved issues \cite{Zhao2024PINNReview}.  

At the same time, a growing body of recent works demonstrates that PINNs are being extended beyond the traditional laminar or subsonic limit into more challenging flow regimes. For example, Jagtap \emph{et al.} \cite{Jagtap2022} applied PINN to solve inverse problems for supersonic compressible flows involving sharps shocks and expansion waves — reconstructing density, pressure and velocity fields from limited boundary and Schlieren-data \cite{Jagtap2022}. On the turbulence side, Eivazi \emph{et al.} showed that PINNs can tackle incompressible turbulent flows by solving the Reynolds-averaged Navier–Stokes (RANS) equations directly — recovering mean-flow and Reynolds-stress fields over boundary layers, airfoil flows and complex terrains without invoking closure models \cite{Eivazi2022}. To tackle the difficulties associated with discontinuities and steep gradients (e.g. shocks), other studies introduced stabilization techniques such as adaptive localized artificial viscosity within the PINN loss\cite{Coutinho2022}. More recently, PINNs have also begun to address complex geometries relevant to realistic engineering applications, pointing toward feasibility for near-real-world domains \cite{Botarelli2025}.  

To address both optimisation stiffness and the prohibitive cost of cold-starting each geometry, we introduce a \textbf{boundary-moving PINN} workflow that generalises the boundary-progressive concept from prior work\cite{GAPINN2022} to a \emph{continuous morphing path} sweeping through an entire NACA65 design family: (i) the airfoil contour is morphed in small Bézier-control-point increments, effectively reusing converged weights of shape \(i\) as warm start for shape \(i+1\); (ii) a geometry-aware dynamic-loss scheduler boosts residual penalties only in the newly exposed high-curvature zones, preventing gradient dilution. 

In a particularly challenging case (65–1010 at \(\alpha = 5^\circ\)), the PINN matches the CFD reference with a \emph{mean relative density error} of \(0.64\% \pm 0.79\%\) and a \emph{mean relative pressure error} of \(0.82\% \pm 1.20\%\), where the values after “\(\pm\)” denote one standard deviation over the computational domain. The corresponding \emph{mean mach number errors} are \(2.60\%\ \pm 2.99\%\,\)  (again reported as mean $\pm$ standard deviation). Across six CFD-validated cases, we typically observe \emph{mean relative errors} in the pressure coefficient distribution \(C_p\) of \(1\text{--}5\%\), with worst-case lift deviations up to \(\sim 10\%\) under the most demanding off-design conditions.

On a single A5000 GPU, per-shape training required approximately \(20\text{--}30\) min; weight reuse along the morphing path reduces wall-clock time by a factor of 3 to 5 compared to cold-start PINNs. Compared with rerunning all operating points using the in-house \textit{SharC} finite-volume solver (with \(10^6\) cells, second-order Euler) \cite{lea2025detached}, the proposed PINN workflow offers substantial throughput gains for family-wide blade screening.

In summary, although PINNs continue to exhibit limitations in robustness, stability and generalisation—especially for high-gradient, turbulent, or off-design turbomachinery flows across multiple operating conditions—recent advances in PINN methodology and applications \cite{PINNReview2024, PINNReviewSparse2025, Zhao2024PINNReview} combined with our proposed boundary-moving workflow have the potential to bring PINNS closer to becoming practical surrogate models for industrial-scale, multi-geometry, multi-operating-condition turbomachinery design.

%======================================================================
%======================================================================
%======================================================================
\section{Methodology}\label{sec:method}

\subsection{Physics-informed neural networks (PINNs)}
Physics-informed neural networks (PINNs)\cite{Raissi2019} are neural networks trained to approximate the solution of partial differential equations (PDEs) by embedding the governing physical laws directly into the training process. Instead of discretizing the domain with a mesh, the solution field $\mathbf{u}(\mathbf{x})$ is represented as a neural network $\mathbf{u}_\theta(\mathbf{x}) = \mathcal{N}_\theta(\mathbf{x})$, which is optimized to satisfy both the PDE residuals in the interior domain $\Omega \subset \mathbb{R}^2$ and the prescribed boundary conditions on $\partial \Omega$. For a system of $n_f$ PDEs with $n_b$ boundary constraints, the governing relations can be written as
\begin{align}
\mathcal{F}_i[\mathbf{u}](\mathbf{x}) &= 0, \quad \mathbf{x}\in \Omega, \quad i=1,\dots,n_f, \\
\mathcal{B}_j[\mathbf{u}](\mathbf{x}) &= 0, \quad \mathbf{x}\in \partial\Omega, \quad j=1,\dots,n_b,
\end{align}
where $\mathcal{F}_i$ denotes the $i$-th PDE residual and $\mathcal{B}_j$ enforces the $j$-th boundary condition. The network parameters $\theta$ are optimized by minimizing a composite loss function
\begin{align}
L(\theta) &= L_\mathrm{PDE} + L_\mathrm{BC}, \\
L_\mathrm{PDE} &= \frac{1}{N_e} \sum_{n=1}^{N_e} \sum_{i=1}^{n_f} 
\bigl| R_i(\mathbf{x}_n;\theta) \bigr|^2, \\
L_\mathrm{BC} &= \frac{1}{N_b} \sum_{m=1}^{N_b} 
\bigl| \mathbf{u}_\theta(\mathbf{x}_m) - \mathbf{u}_\mathrm{BC}(\mathbf{x}_m) \bigr|^2,
\end{align}
where $R_i(\mathbf{x}_n;\theta)$ denotes the PDE residual at collocation point $\mathbf{x}_n$, and $\mathbf{u}_\mathrm{BC}$ represents the target boundary values. Collocation points $\{\mathbf{x}_n\}$ and $\{\mathbf{x}_m\}$ are distributed throughout the interior and along the boundaries of the domain, ensuring that the network learns a continuous approximation that satisfies the physical laws everywhere in $\Omega$. Automatic differentiation \cite{paszke2017automatic} computes the derivatives appearing in the PDE residuals directly from the network outputs, allowing accurate evaluation without numerical discretization. Once trained, $\mathbf{u}_\theta(\mathbf{x})$ provides a continuous, mesh-free representation of the solution field that can be queried at arbitrary points, even where no collocation points were sampled. This approach is particularly advantageous for problems involving complex geometries, parameterized domains, or coupled multi-physics, as it naturally accommodates arbitrary shapes and continuous variations in operating conditions. However, training PINNs can be challenging due to stiff interactions between PDE residual and boundary-condition losses, especially near boundaries or in high-gradient regions, and convergence is often sensitive to network initialization, collocation point placement, and problem complexity. These limitations motivate the development of geometry-aware weighting and boundary-progressive training strategies, which are described in Sections~\ref{sec:relax} and~\ref{sec:weight}, to improve stability, accelerate convergence, and enhance prediction accuracy for industrial-scale turbomachinery applications.

\subsection{Baseline PINN formulation}\label{sec:pinn21}
%----------------------------------------------------------------------

First, we define the governing equations. The mathematical model consists of the two-dimensional, Euler equations in conservative form:
%--------------------------- Governing PDE ----------------------------

\begin{equation}
  \partial_{x}
    \begin{pmatrix}
      \rho u \\
      \rho u^{2}+p \\
      \rho u v \\
      \rho u H
    \end{pmatrix}
 +\partial_{y}
    \begin{pmatrix}
      \rho v \\
      \rho u v \\
      \rho v^{2}+p \\
      \rho v H
    \end{pmatrix}=0,
  \label{eq:euler2d_cons}
\end{equation}
where \(H=E+p/\rho\) and
\(E=e+\tfrac12(u^{2}+v^{2})\).
and the thermal state equation for an ideal gas is given by
\(p=(\gamma-1)\rho e,\;\gamma=1.4\).

In PINNs, large differences in magnitudes among output variables tend to cause instability and degrade accuracy. To mitigate this, we adopt the non-dimensional form of the equations to render all relevant terms  $\mathcal{O}(1)$. Specifically, we define:

\[
\begin{aligned}
& (x,y) \rightarrow (\tilde x, \tilde y) = \frac{(x,y)}{L_{\mathrm{ref}}}, \\
& \rho \rightarrow \tilde \rho = \frac{\rho}{\rho_{\mathrm{ref}}}, 
\quad p \rightarrow \tilde p = \frac{p}{p_{\mathrm{ref}}}, \\
& \boldsymbol v \rightarrow \tilde{\boldsymbol v} = \frac{\boldsymbol v}{a_{\mathrm{in}}}, 
\quad \rho_{\mathrm{ref}} = \frac{p_{\mathrm{ref}}}{a_{\mathrm{in}}^2}.
\end{aligned}
\]

Hence each variable is scaled to $\mathcal{O}(1)$, and the Mach number enters only through $\tilde{\boldsymbol v}$ (i.e.\ the dimensionless velocity), improving training stability and accuracy.

%-------------------- Primitive-variable formulation ------------------

Combining \eqref{eq:euler2d_cons} with the ideal-gas equation of state, 
the Euler equations can be expressed in primitive variables as
\begin{equation}\label{eq:prim2d}
\begin{aligned}
 & u\,\partial_{x}\rho
  + v\,\partial_{y}\rho
  + \rho\left(\partial_{x}u + \partial_{y}v\right) = 0, \\[2pt]
 & u\,\partial_{x}u
  + v\,\partial_{y}u
  + \rho^{-1}\,\partial_{x}p = 0, \\[2pt]
 & u\,\partial_{x}v
  + v\,\partial_{y}v
  + \rho^{-1}\,\partial_{y}p = 0, \\[2pt]
 & u\,\partial_{x}p
  + v\,\partial_{y}p
  + \gamma p\left(\partial_{x}u + \partial_{y}v\right) = 0.
\end{aligned}
\end{equation}

This formulation avoids stiff density-division terms during 
automatic differentiation.

%--------------------- Inlet total-to-static map ----------------------

Also, for inviscid adiabatic flow, we have:

\begin{equation}
  \check{T}_{\text{in}}
    = 1-\frac{\gamma-1}{2}\,
        \check{\boldsymbol v}\!\cdot\!\check{\boldsymbol v},\qquad
  \check{\rho}_{\text{in}}
    = \gamma\,\check{T}_{\text{in}}^{\,1/(\gamma-1)},\qquad
  \check{p}_{\text{in}}
    = \check{T}_{\text{in}}^{\,(\gamma-1)/\gamma},
  \label{eq:tot2stat}
\end{equation}
giving the static targets used in the inlet MSE term \\[5pt]

%------------------------ Network definition --------------------------
\paragraph{Network architecture and positivity enforcement}

The network $\mathcal{N}_\theta$ is chosen as a fully connected (MLP) model with layer sizes $(2,20,20,20,4)$ and tanh activation functions to aid in capturing smooth variations and avoid large output magnitude disparities\cite{post_investigation_2022}. 

It predicts the vector
\[
[\check{\rho},\; \check{p},\; \check{u},\; \check{v}]^{\!\top}.
\]
To guarantee strictly positive values for density and pressure (i.e.\ $\check{\rho}, \check{p} > 0$), we enforce an exponential mapping:
\begin{equation}\label{eq:posmap_opt}
  \check{\rho} = \exp\!\left(N_\rho\right), \qquad
  \check{p} = \exp\!\left(N_p\right),
\end{equation}
where $N_\rho$, $N_p$ are (scalar) outputs of $\mathcal{N}_\theta$.  
This follows common practice in Euler-PINN literature for preserving physical consistency and preventing instability from negative or zero density/pressure\cite{post_investigation_2022}.

%------------------------------ Losses --------------------------------
\paragraph{Interior (PDE) loss}
The losses function for the Partial Differential Equation(PDE) part are defined as:
\[
  L_{\mathrm{PDE}}
  =\frac1{N_e}\sum_{n=1}^{N_e}\sum_{i=1}^{4}
    \bigl|R_i(\mathbf x_n;\theta)\bigr|^{2}.
\]

\paragraph{Boundary losses}
\begin{align}
  L_{\text{in}}  &=
     \text{MSE}\bigl[\check{p}-\check{p}_{\text{in}}\bigr]
    +\text{MSE}\bigl[\check{\rho}-\check{\rho}_{\text{in}}\bigr]
    +\text{MSE}\bigl[\check{u}-|\check{u}|\bigr], \\[2pt]
  L_{\text{out}} &=
     \text{MSE}\bigl[\check{p}-\check{p}_{\text{out}}\bigr]
    +\text{MSE}\bigl[\check{u}-|\check{u}|\bigr], \\[2pt]
  L_{\text{wall}}&=
     \text{MSE}\bigl[(\check{u},\check{v})\!\cdot\!\hat{\boldsymbol n}\bigr], \\[2pt]
  L_{\text{per}} &=
     \text{MSE}\bigl[\mathcal N_\theta(x,y)-\mathcal N_\theta(x+x_p,y)\bigr],
\end{align}
with \(L_{\mathrm{BC}}=L_{\text{in}}+L_{\text{out}}+ L_{\text{wall}}+L_{\text{per}}\).

\paragraph{Total loss (fixed weights)}
\begin{equation}
  L
  =\omega_{\mathrm{PDE}}L_{\mathrm{PDE}}
  +\omega_{\mathrm{BC}}L_{\mathrm{BC}},\qquad
  \omega_{\mathrm{PDE}}=1,\;
  \omega_{\mathrm{BC}}=10.
  \label{eq:totloss}
\end{equation}

%----------------------------------------------------------------------
\subsection{Problem defination and geometry setup}
\label{sec:geom}
%----------------------------------------------------------------------

\paragraph{Airfoil family and nomenclature}
All test cases employ \textbf{NACA 6-series} profiles, whose code
\texttt{65-$m$$t$} follows the convention  
$6$ (series) – $5$ (pressure–recovery factor $\ell = 0.5$) – $\!m$
(design lift in 0.1 increments) – $\!t$ (thickness in \% chord) \cite{NACA6Series}.\footnote{%
For example, NACA 65-0010 is symmetrical ($m=0$) with $10\%$ thickness,
whereas 65-0610 has $C_{L,\mathrm{design}}=0.6$ and the same thickness
distribution.}  
The baseline set comprises  
\(\text{65-0005},\;0010,\;0012\) and  
three cambered extensions for each thick profile:
\(\text{65-0310},\;0610,\;1010\) 

\paragraph{Flow conditions}
All simulations assume a calorically and thermally perfect gas
($\gamma=1.4,\,R=\SI{287}{J\,kg^{-1}K^{-1}}$) .  
Inlet total conditions are fixed to  
\(p_{t,\mathrm{in}} = \SI{1}{\bar}\) and
\(T_{t,\mathrm{in}} = \SI{300}{\kelvin}\), while outlet static pressure
takes two levels,
\(p_{\mathrm{out}} = \SI{0.95}{\bar},\;\SI{0.80}{\bar}\).
Each shape is analysed at two incidences,
\(\alpha = 0^{\circ},\;5^{\circ}\).

\begin{figure}[H]
  \centering
  \includegraphics[width=1\linewidth]{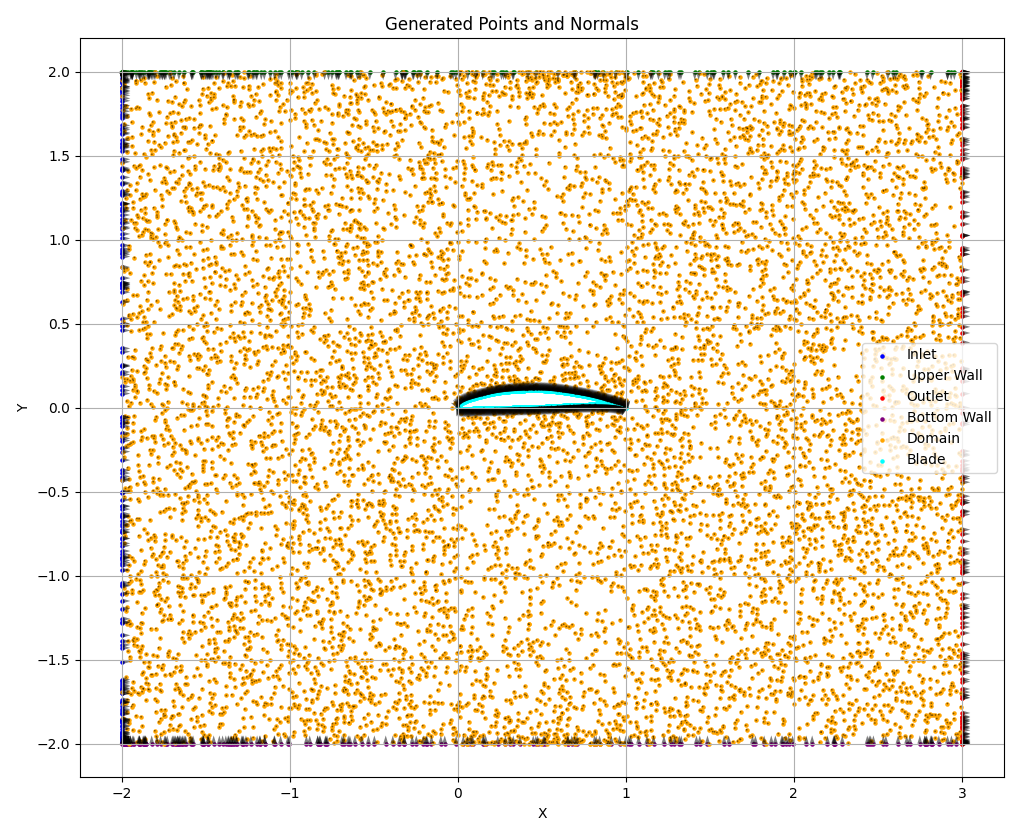}
  \caption{Initial collocation layout for NACA 65-1010.  Orange dots are
           interior Sobol points; coloured dots match the tags in
           Table~\ref{tab:sampling}.}
  \label{fig:initial_points}
\end{figure}

\paragraph{Collocation budget and placement}
The PINN uses a total of \num{4000} interior collocation points, which are drawn from a Sobol quasi-random (low-discrepancy) sequence and then partitioned equally across the front, mid, and rear thirds of the domain, supplemented by at least \num{1000} boundary points per airfoil (Tab.~\ref{tab:sampling}).  Sobol sequences are a class of low-discrepancy, quasi-random point sets that are specifically designed to minimise clustering and gaps in multi-dimensional spaces compared with purely random sampling, leading to more uniform coverage of the domain and improved space-filling properties for surrogate models and numerical integration \cite{Sobol1967}.  This balanced, low-discrepancy sampling strategy helps ensure that both interior flow features and boundary conditions are represented more evenly in the training dataset, which is critical for stable PINN training.  Fig.~\ref{fig:initial_points} illustrates the initial point layout for the 65-1010 profile.

%------------------------  Figure (fixed position) --------------------

%------------------------ Collocation table ---------------------------
\begin{table}[H]
  \centering
  \caption{Collocation-point budget and placement rules
           (per airfoil).\label{tab:sampling}}
  \resizebox{0.95\linewidth}{!}{
  \begin{tabular}{@{}l c l p{6cm}@{}}
    \toprule
    \textbf{Tag / region} & \textbf{Pts} & \textbf{Placement rule} & \textbf{Notes} \\ \midrule
    Interior domain       & \num{4000} & Split $1{:}1{:}1$ (front / mid / rear) & Sobol in each rectangle \\[2pt]
    Upper slip wall$^{\dagger}$ & \num{300} & Uniform arc-length & Includes suction side of blade \\[2pt]
    Lower slip wall$^{\dagger}$ & \num{300} & Uniform arc-length & Includes pressure side of blade \\[2pt]
    Inlet plane           & \num{200} & Random on $y\!\in[-1,1]$ & Stagnation inflow via Eq.\,\eqref{eq:tot2stat} \\[2pt]
    Outlet plane          & \num{200} & Random on $y\!\in[-1,1]$ & Static-pressure outlet \\[2pt]
    Upper blade surface   & \num{300} & Equidistant along contour & Periodic pair with lower blade \\[2pt]
    Lower blade surface   & \num{300} & Equidistant along contour & Periodic pair with upper blade \\[2pt]\midrule
    \textbf{Total}        & \textbf{\num{5400}} & & \\ \bottomrule
  \end{tabular}
  }
  \vspace{4pt}\footnotesize
  $^{\dagger}$Straight channel walls at $y=\pm1$ are merged with the
  respective blade sides, yielding a continuous slip boundary for the PINN.
\end{table}

%----------------------  Test-matrix (56 runs) ------------------------

\begin{table}[H]
  \centering
  \caption{Design matrix. All 20 combinations are solved with PINNs; the 12 starred (\cfdmark)  
           cases are additionally simulated with the CFD reference solver.}
  \label{tab:test_matrix}
  \begin{tabular}{@{}lcccc@{}}
    \toprule
    \textbf{Airfoil} &
    $(0^{\circ},\,0.95)$ &
    $(0^{\circ},\,0.80)$ &
    $(5^{\circ},\,0.95)$ &
    $(5^{\circ},\,0.80)$ \\ \midrule
    65-0005 & \cmark   & \cmark & \cmark & \cmark   \\
    65-0010 & \cfdmark & \cmark & \cmark & \cfdmark \\[2pt]
    65-0310 & \cmark   & \cfdmark & \cmark & \cmark   \\
    65-0610 & \cfdmark & \cmark & \cfdmark & \cmark   \\
    65-1010 &          &        & \cmark & \cfdmark \\[2pt]
    \textbf{Totals} &
      3\,\cfdmark / 14\,\cmark &
      2\,\cfdmark / 14\,\cmark &
      2\,\cfdmark / 14\,\cmark &
      3\,\cfdmark / 14\,\cmark \\ \bottomrule
  \end{tabular}

  \vspace{4pt}
  \footnotesize
  \cfdmark\;=\;PINN\,+\,CFD\quad\quad
  \cmark\;=\;PINN only
\end{table}

%----------------------------------------------------------------------
\subsection{Boundary-progressive convergence strategy}\label{sec:relax}
%======================================================================

Complex camber lines and strong pressure gradients notoriously hamper
PINN optimisation: PDE residuals and boundary-condition Minium Square Error(MSE) terms
compete, driving the loss toward poor local minima \cite{Krishnapriyan2021,Rathore2024}.Hard-encoded boundary enforcement methods, such as constructing special trial functions that exactly satisfy prescribed conditions, have been explored in PINN research \cite{Sukumar2021ExactBC}, but they are inherently difficult to design for and generalise to geometrically intricate boundaries like complex blade walls \cite{Sukumar2021ExactBC}.

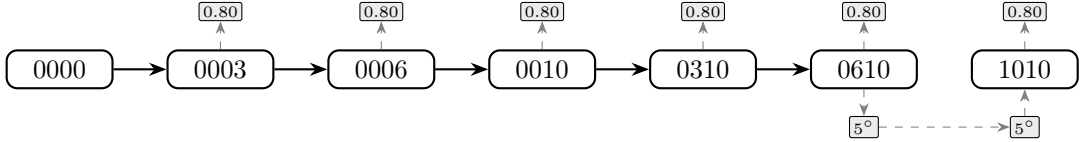
\begin{figure}[H]
  \centering
\begin{tikzpicture}[
  >=Stealth,
  node distance = 18mm and 7mm,
  main/.style = {draw, rounded corners, thick, minimum width=14mm, font=\footnotesize},
  tiny/.style = {draw, rounded corners=0.8pt, inner sep=1.5pt, fill=gray!15, font=\tiny},
  warm/.style = {->, thick},
  bc/.style   = {->, dashed, gray, thin}
]

% ---------------- main chain ----------------
\node[main] (n0000) {0000};
\node[main, right=of n0000] (n0003) {0003};
\node[main, right=of n0003] (n0006) {0006};
\node[main, right=of n0006] (n0010) {0010};
\node[main, right=of n0010] (n0310) {0310};
\node[main, right=of n0310] (n0610) {0610};
\node[main, right=of n0610] (n1010) {1010};

% ---------- warm-start arrows (horizontal) ----------
\draw[warm] (n0000) -- (n0003);
\draw[warm] (n0003) -- (n0006);
\draw[warm] (n0006) -- (n0010);
\draw[warm] (n0010) -- (n0310);
\draw[warm] (n0310) -- (n0610);

% 注意这里不直接从 0610 连到 1010
% 而是通过下面的 “5°” 节点
% ---------- boundary branches: 0.80 bar above; 5° below ----------
\foreach \n in {n0003,n0006,n0010,n0310,n0610,n1010}{
  % 0.80 bar (above)
  \path (\n.north) ++(0,0.5cm) node[tiny] (\n low) {0.80};
  \draw[bc] (\n.north) -- (\n low.south);
}

% ---------- special branch from 0610 down to its 5° node, then over to 1010's 5° node ----------
% create the 5° nodes under 0610 and 1010
\path (n0610.south) ++(0,-0.5cm) node[tiny] (d0610) {$5^{\circ}$};
\path (n1010.south) ++(0,-0.5cm) node[tiny] (d1010) {$5^{\circ}$};

% vertical branch down from 0610 to its 5° node
\draw[bc] (n0610.south) -- (d0610.north);

% horizontal connection from 0610's 5° node to 1010's 5° node
\draw[bc] (d0610.east) -- (d1010.west);

% then vertical up from 1010's 5° node to 1010 (if needed visual arrow)
\draw[bc] (d1010.north) -- (n1010.south);

\end{tikzpicture}
\caption{Main chain with special branch: 0610 to 1010 via 5° node.}\label{fig:main_chain}
\end{figure}

To mitigate these challenges, we adopt a progressive “easy-to-hard” training schedule. Central to this approach is the concept of a \emph{main chain} of airfoil geometries, which defines a sequence of profiles where each new variant is initialised using the converged weights of its predecessor. This allows knowledge from simpler configurations to be propagated to more complex shapes, accelerating convergence and stabilising the optimisation. Figure~\ref{fig:main_chain} illustrates this main chain and shows how branches can be added to handle alternative boundary conditions such as different outlet pressures or incidence angles.  

\begin{enumerate}[label=(\roman*)]
  \item \textbf{Stepwise boundary switch.}\;
        We begin with a pure channel flow
        (\emph{inviscid Riemann inflow / outflow, no blade}).  
        After full convergence, we insert the blade and apply the true
        static-pressure outlet, then retrain until the new configuration
        converges.
  \item \textbf{Geometry warm-starting.}\;
        The converged weights of each geometry initialise the next:
        65-0000 → 65-0005 → 65-0010 .
        Each cambered family is seeded from its zero-camber (or thinner)
        parent; the two warm-start paths are compared in Section \ref{sec:results}.
\end{enumerate}

\paragraph{Training itinerary}

\begin{center}\footnotesize
\begin{tabular}{@{}l l@{}}
\toprule
\textbf{Stage} & \textbf{Action (\cmark = train; BC switch after convergence)}\\\midrule
1 & 65-0000 \cmark\;$\longrightarrow$ weights\,$\mapsto$ 65-0003\\[2pt]
3 & 65-0005 \cmark\;$\longrightarrow$ weights\,$\mapsto$ 65-0010\\[4pt]
4 & (seed 0010)\;$\longrightarrow$ 65-0310 \cmark\;$\longrightarrow$ 65-0610 \cmark\;$\longrightarrow$ 65-1010 \cmark\\
\bottomrule
\end{tabular}
\end{center}
\normalsize

The proposed progressive schedule proves effective for several intertwined reasons.  First, it alleviates the common conflict between enforcing PDE residuals and imposing boundary conditions: by initially solving a simple hyperbolic Riemann problem — which omits stiff static-pressure constraints and complex wall geometries — the network establishes a consistent base flow in an easy setting.  Only once this baseline is achieved do we introduce the blade geometry and outlet pressure; at that stage, the required adjustments are local rather than global, thereby avoiding abrupt jumps that often destabilize training.  Moreover, the converged weights from the baseline run implicitly encode the potential-flow topology.  As a result, when the geometry is perturbed slightly (e.g., modest changes in camber) or loading conditions vary marginally, fine-tuning from this warm start converges 3–5 times faster than a cold-start initialization, while substantially reducing the risk of settling into non-physical minima.  Finally, this strategy scales efficiently over entire airfoil families: because each new variant is warm-started from a closely related case, evaluating ten additional shapes typically requires only about $20 \% $ of the GPU time that would be needed if each case were trained independently.  Section \ref{sec:results} presents a quantitative account of this speed-up.

%----------------------------------------------------------------------
\subsection{Geometry-aware dynamic loss weighting}\label{sec:weight}
%======================================================================

Standard PINNs often diverge or stagnate once the blade is inserted:
domain-residual and boundary MSE terms pull the optimiser in opposite
directions, producing “saw-tooth” loss curves and non-physical plateaus
\cite{Krishnapriyan2021,Rathore2024}.  Recent studies therefore embed
\emph{geometry information} directly into the loss
\cite{GAPINN2022}.  Our implementation follows this line and introduces a
smooth weight field \(w(\mathbf x)\) that amplifies residuals
near curved walls and inlet/outlet planes while leaving the core flow
largely untouched:

\[
L_{\text{PDE}}^{\ast}
   = \frac{1}{N_e}\sum_{n=1}^{N_e}\sum_{i=1}^{4}
     \bigl[w(\mathbf x_n)\,R_i(\mathbf x_n;\theta)\bigr]^2,
\qquad
w:\Omega\subset\mathbb R^2\rightarrow\mathbb R_{>0}.
\tag{17}
\]

\paragraph{Weight definition}

Let \(D_j(\mathbf x)\) be the signed Euclidean distance from point
\(\mathbf x\) to the $j$-th boundary segment  
(\emph{inlet}, \emph{outlet}, \emph{upper} and \emph{lower} walls,
where the straight walls morph to the true blade contour inside
\(x\in[0,1]\)).  
We assign

\[
w(\mathbf x)=
1+\bigl(w_{\mathrm{bd}}-1\bigr)\!
  \sum_{j=1}^{4}\exp\!\Bigl[-D_j(\mathbf x)/d\Bigr],
\quad
d=0.20,\;w_{\mathrm{bd}}=60,
\tag{18}
\]

so \(w\approx w_{\mathrm{bd}}\) on the surface and smoothly decays to 1
within \(10\%\) chord length.

\begin{figure}[H]
  \centering
  \includegraphics[width=1\linewidth]{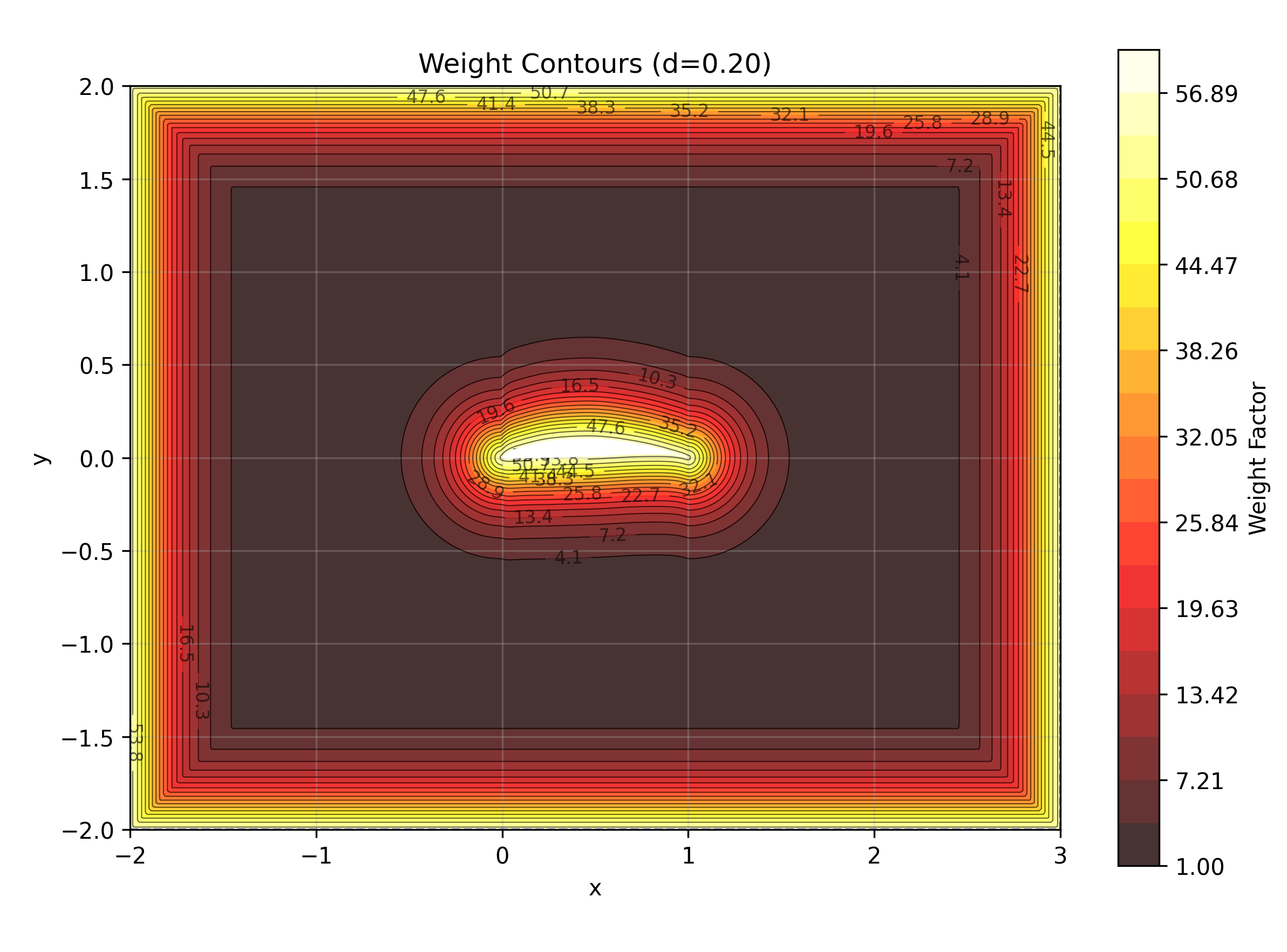}
  \caption{Spatial distribution of the geometry-aware weight field
           (case 65-1010).  Hottest zones hug the blade and channel
           walls; the core region remains nearly unweighted.}
  \label{fig:weight_field}
\end{figure}

Near the blade surface we simultaneously face the strongest physics–geometry coupling \emph{and} the highest engineering interest, so we deliberately amplify the residual there: the larger weight forces the optimiser to honour boundary conditions first, then propagate corrections into the core flow, eliminating the usual PDE–BC tug-of-war while concentrating accuracy where the design is most sensitive; although the extra multiplications add a few seconds per epoch, the weight field is fixed and the overhead is negligible relative to the total runtime.

%------------------------ Network & training --------------------------
\paragraph{Training}

The training consists of an initial phase using the Adam optimizer for 50 steps with a learning rate of \(\eta = 10^{-2}\), which allows the network to transition from the simple Riemann flow to the baseline NACA 0006 configuration. This is followed by approximately 3000 iterations of L-BFGS optimization, continuing until the loss changes by less than \(10^{-4}\). After each update of the boundary conditions or geometry, additional L-BFGS iterations are performed to refine the solution: typically, 1000 iterations are used for most cases, while the more challenging NACA 65-0610 and NACA 65-1010 profiles at an incidence of \(\alpha = 5^\circ\) require 2000 and 4000 iterations, respectively, to achieve the desired accuracy due to the increased complexity near the blade surfaces. This staged training strategy ensures stable convergence, efficiently propagates information from simpler configurations to more complex geometries, and maintains high fidelity in regions of steep gradients or strong boundary-layer interactions.

%======================================================================
\section{Results}\label{sec:results}
%======================================================================

\subsection{Convergence behaviour}\label{sec:res_conv}
\begin{figure}[H]
  \centering
  \setlength{\tabcolsep}{4pt}
  \renewcommand{\arraystretch}{1.0}

  \makebox[\textwidth][c]{%
    \begin{tabular}{ccc}
      \subcaptionbox*{(a)\;\;Cold start}{%
        \includegraphics[height=4cm,keepaspectratio]{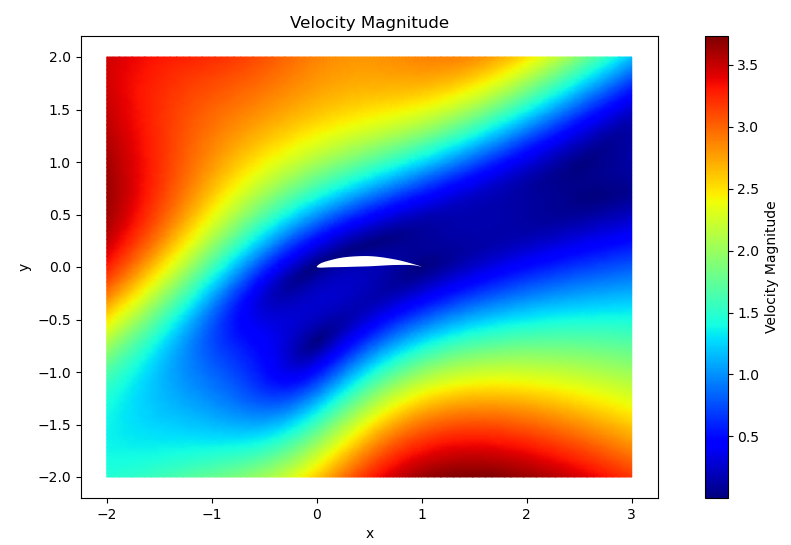}} &
      \subcaptionbox*{(b)\;\;Warm-start only}{%
        \includegraphics[height=4cm,keepaspectratio]{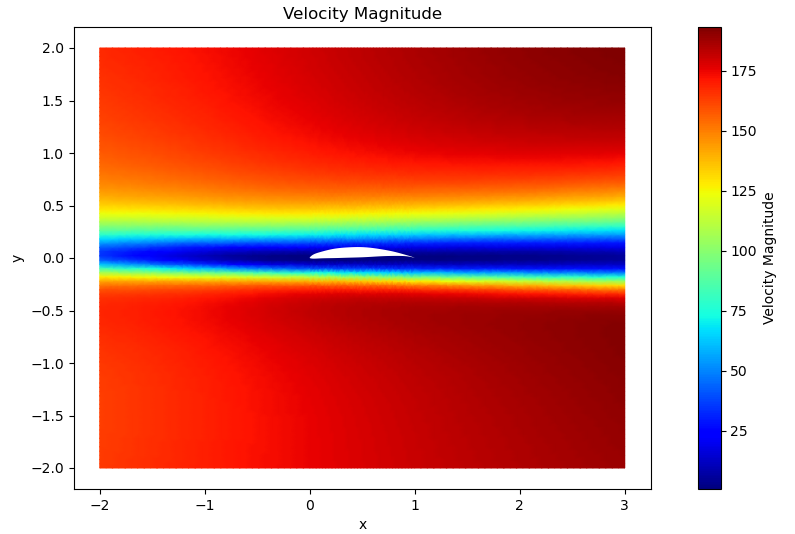}} &
      \subcaptionbox*{(c)\;\;Combined}{%
        \includegraphics[height=4cm,keepaspectratio]{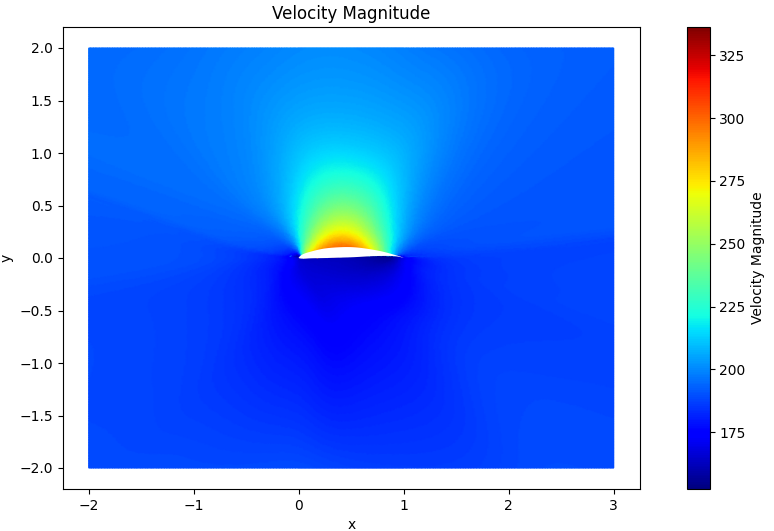}}
    \end{tabular}%
  }

  \caption{Velocity contours for the three training strategies.
           Cold start is non-physical; warm-start only under-predicts
           near the blade; combined schedule matches CFD closely.}
  \label{fig:mach_triptych}
\end{figure}

From Fig.~\ref{fig:mach_triptych} (and the loss curves in Fig.~\ref{fig:loss_history}),
the \emph{cold-start} PINN collapses to a non-physical near-stagnant state, with
velocity magnitudes approaching zero across the domain. This behaviour is consistent
with the primitive form \eqref{eq:prim2d}: taking $\check u\!\approx\!0$,
$\check v\!\approx\!0$ and small pressure gradients drives all residuals toward low
values, while the slip (no-penetration) condition
$\hat{\boldsymbol n}\!\cdot\!(\check u,\check v)=0$ on the blade is trivially satisfied.
In other words, the network can reduce both PDE loss and wall-loss by converging to a
degenerate solution.

In the \emph{warm-start only} run, the bulk field is recovered, but a spurious
near-wall band with strongly reduced velocity appears around the blade. This reflects
the persistent conflict between the slip-wall MSE and the interior PDE residual:
the optimiser lowers the total loss by quenching the velocity next to the surface.

The \emph{combined} strategy resolves both issues. Warm-starting constrains the search
away from poor basins, and the geometry-aware weighting increases the penalty of
near-wall residuals, preventing the network from creating a zero-velocity strip to
satisfy the wall condition. As a result, the final prediction remains physical in the
vicinity of the blade and matches the global flow features seen in the reference
solution.

%----------------------------------------------------------------------
\subsection{Accuracy versus SharC}\label{sec:res_acc}
%----------------------------------------------------------------------

\paragraph{Global error metrics}  
Table \ref{tab:global_l2} summarises the \(MSE\) error of density(\(\rho\)), Mach number (\(M\)), and static pressure (\(p\)) over all calculation domine.  Mean errors remain below 1 \%, and the maximum local error never exceeds 20 \%(exclude Mach number because of the nearly zero velocity nearby the leading edge).

\begin{table}[H]
  \centering
  \caption{Global \(MSE\) error: progressive-PINN vs.\ SharC CFD.}
  \label{tab:global_l2}
  \begin{tabular}{@{}lccc@{}}
    \toprule
    & \(C_p\) [\%] & \(M\) [\%] & \(p\) [\%] \\ \midrule
    Mean error      & 0.64 &  2.6 & 0.82 \\
    Standard dev.   & 0.79 & 2.99 & 1.2 \\
    Maximum error   & 16.3 & 36.2 & 17.4\\ \bottomrule
  \end{tabular}
\end{table}

\begin{figure}[H]
  \centering
  \includegraphics[width=1\linewidth]{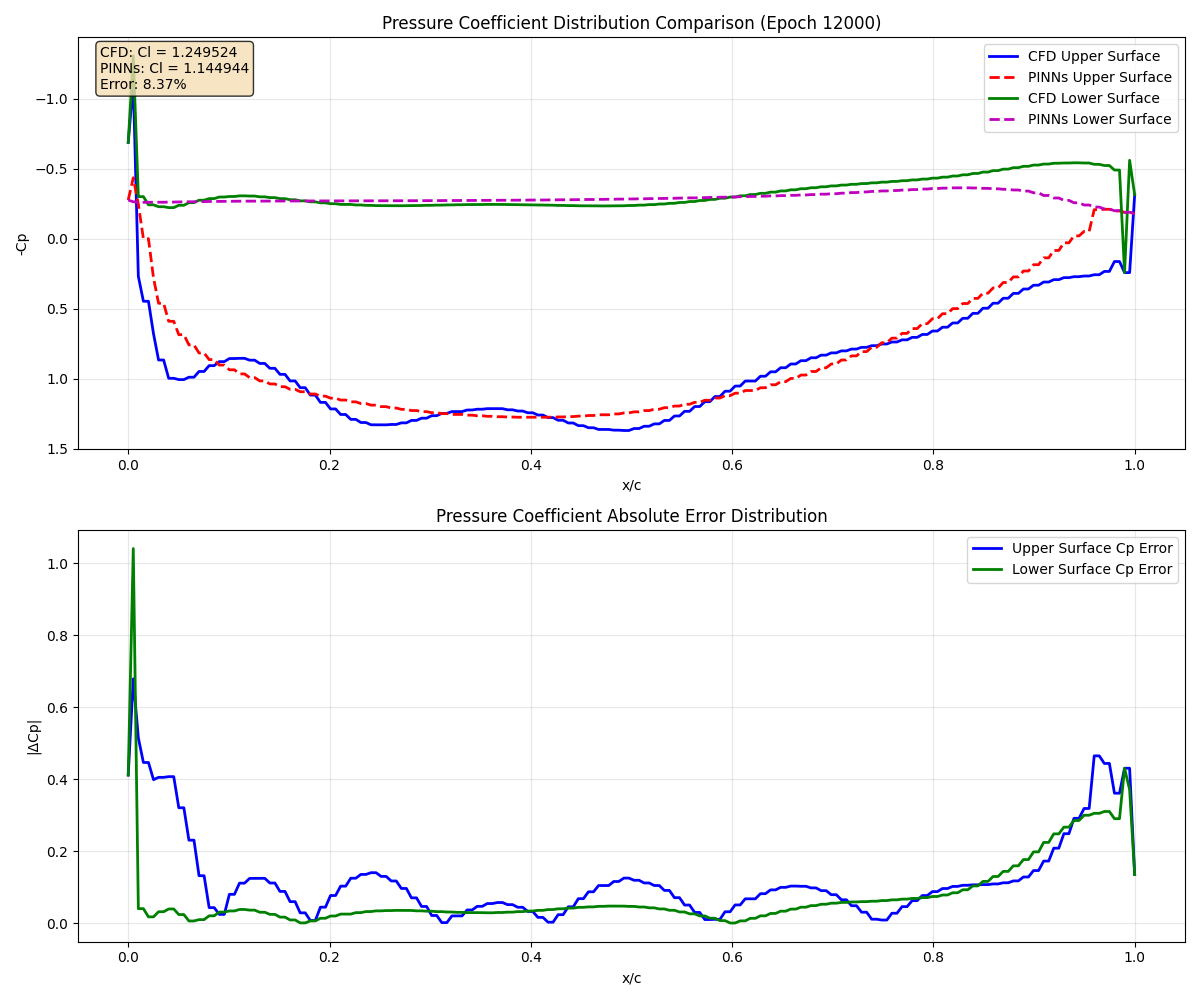}
  \caption{Surface pressure coefficient \(C_p\) for NACA65‑1010
           (\(\alpha = 0^{\circ}\), \(p_{\text{out}} = 0.8\)bar).  
           PINN (solid blue) and CFD (dashed black) for
           pressure side (positive \(C_p\)) and suction side
           (negative \(C_p\)). large errors appear only near
           the leading edge and trailing edge where
           \(|C_p|\) approaches zero.}
  \label{fig:cp_curve_1010}
\end{figure}

\begin{figure}[H]
  \centering
  \includegraphics[width=1\linewidth]{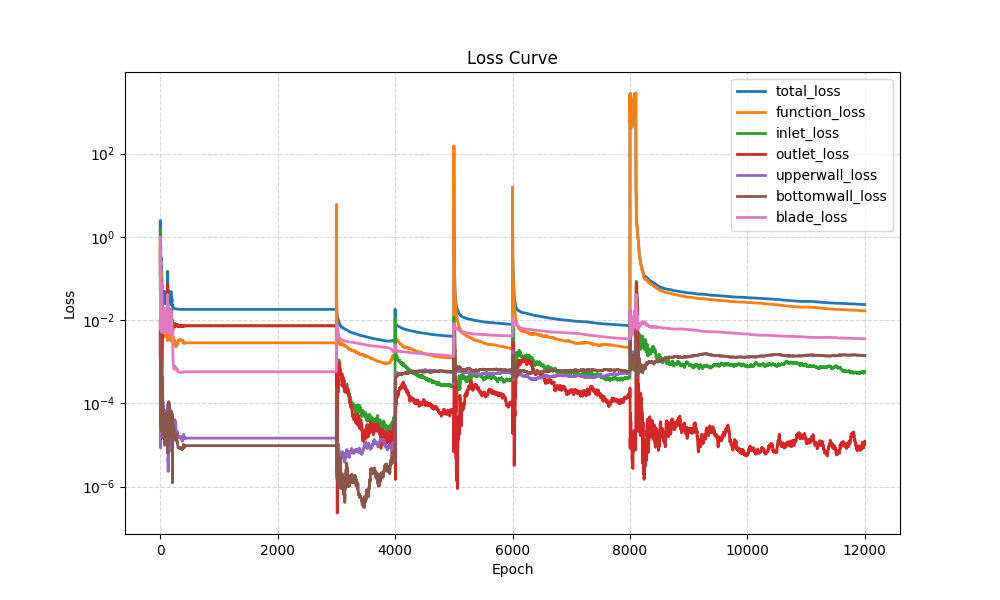}
  \caption{Training loss. Every loss bump refer to a boundary condition restart.}
  \label{fig:loss_history}
\end{figure}

\paragraph{Representative case}  
Figure \ref{fig:loss_history} and \ref{fig:cp_curve_1010} compares surface \(C_p\) and the Density, Pressure, Mach
contours for airfoil 65-1010 at
\(\alpha\!=\!5^{\circ},\;p_{\text{out}}\!=\!0.80\;\text{bar}\).  The
PINN reproduces the suction-side peak and the pressure recovery zone
within 5 \% of the CFD reference.

%----------------------------------------------------------------------
\begin{figure}[H]
  \centering
  \setlength{\tabcolsep}{2pt}% horizontal gap between subfigures
  \renewcommand{\arraystretch}{1.0}% no extra vertical stretch

  % ---------- ROW 1 : PINN -------------------------------------------
  \makebox[\linewidth][c]{%
    \begin{tabular}{@{}ccc@{}}
      \subcaptionbox*{(a)\;\;$\rho$ PINNs}{%
        \includegraphics[height=5cm,keepaspectratio]{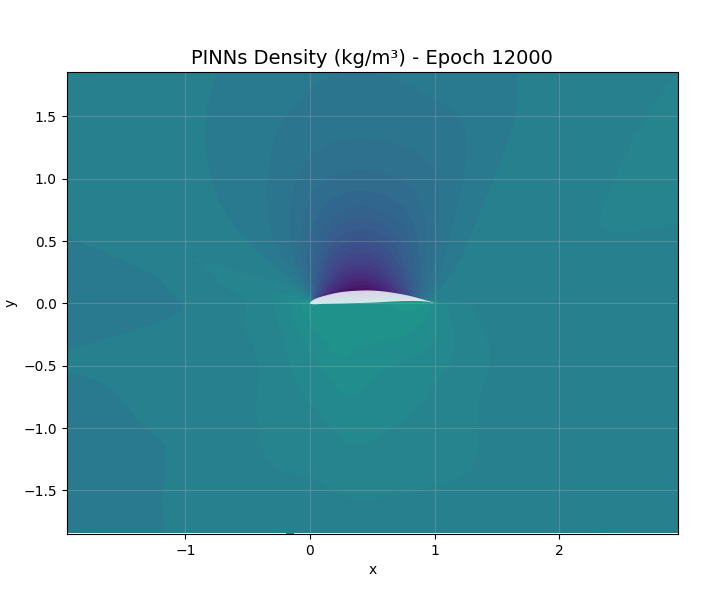}} &
      \subcaptionbox*{(b)\;\;$\rho$ CFD}{%
        \includegraphics[height=5cm,keepaspectratio]{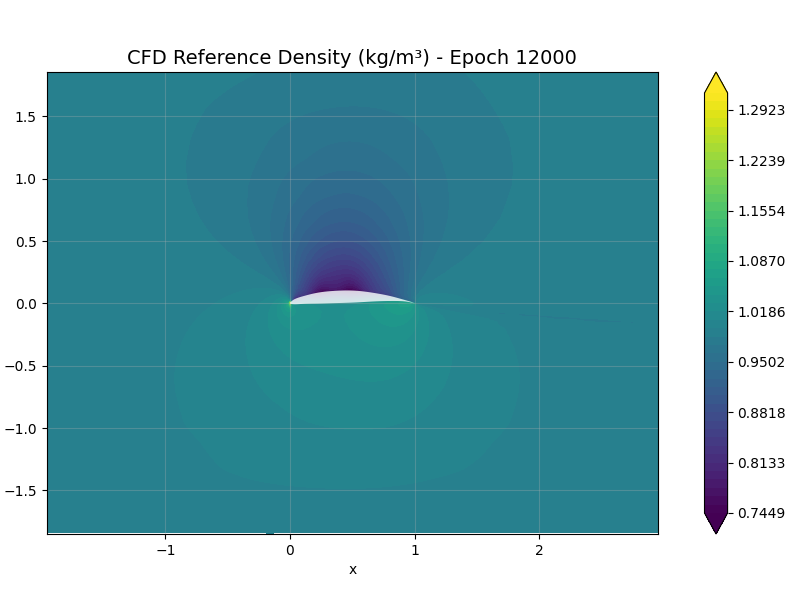}} &
      \subcaptionbox*{(c)\;\;$\rho$ Error}{%
        \includegraphics[height=5cm,keepaspectratio]{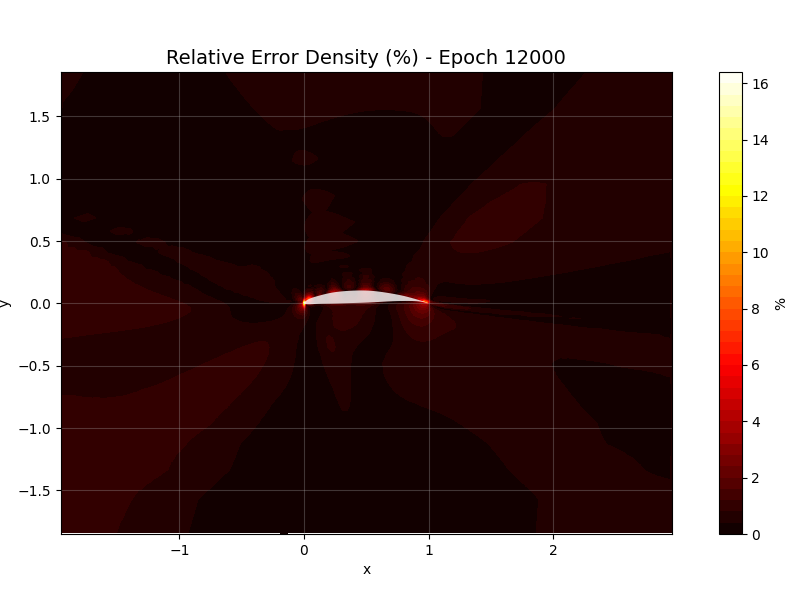}}
    \end{tabular}%
  }

  \vspace{2pt} % small vertical gap

  % ---------- ROW 2 : CFD -------------------------------------------
  \makebox[\linewidth][c]{%
    \begin{tabular}{@{}ccc@{}}
      \subcaptionbox*{(d)\;\;$p$ PINNs}{%
        \includegraphics[height=5cm,keepaspectratio]{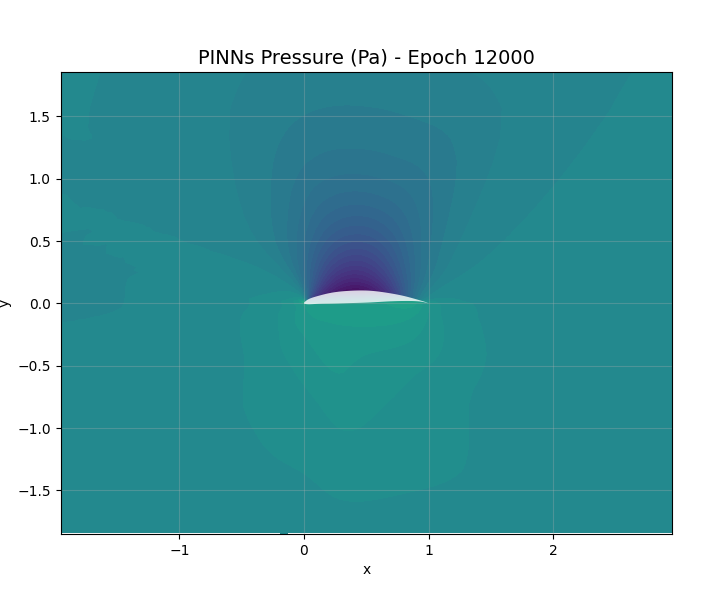}} &
      \subcaptionbox*{(e)\;\;$p$ CFD}{%
        \includegraphics[height=5cm,keepaspectratio]{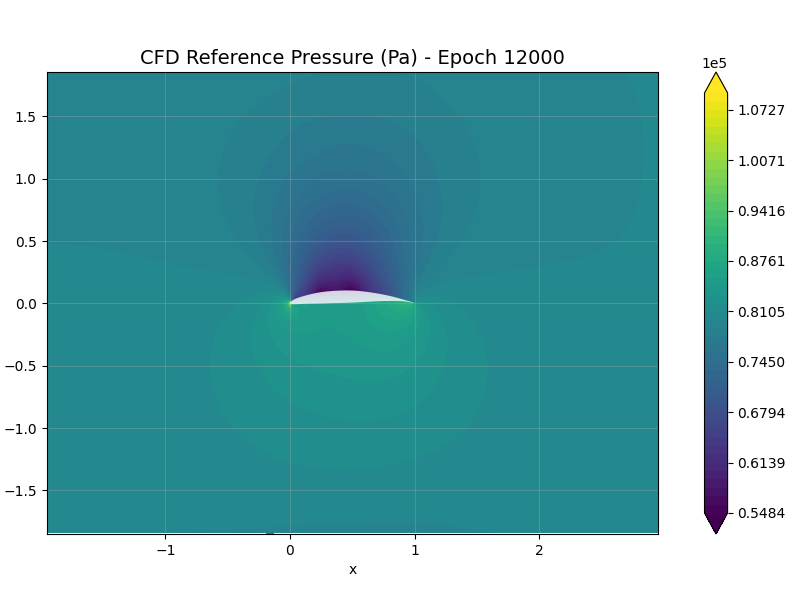}} &
      \subcaptionbox*{(f)\;\;$p$ Error}{%
        \includegraphics[height=5cm,keepaspectratio]{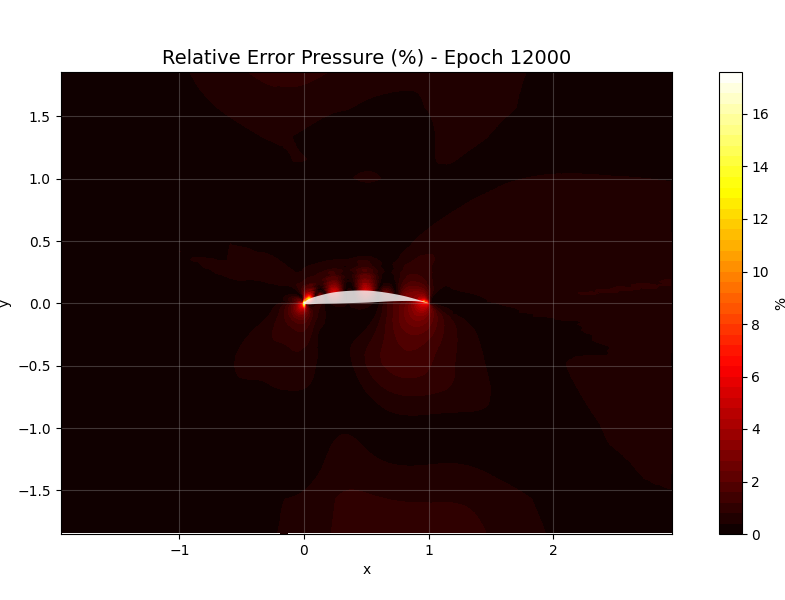}}
    \end{tabular}%
  }

  \vspace{2pt} % small vertical gap

  % ---------- ROW 3 : RELATIVE ERROR --------------------------------
  \makebox[\linewidth][c]{%
    \begin{tabular}{@{}ccc@{}}
      \subcaptionbox*{(g)\;\;$M$ PINNs}{%
        \includegraphics[height=5cm,keepaspectratio]{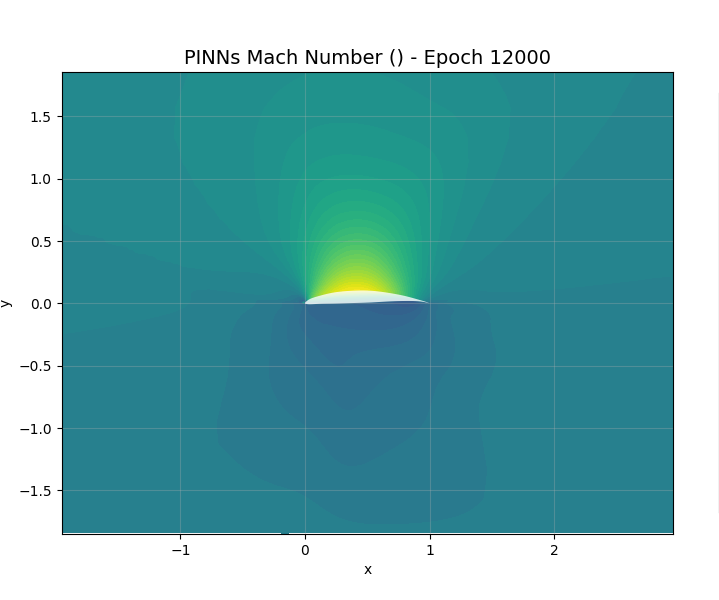}} &
      \subcaptionbox*{(h)\;\;$M$ CFD}{%
        \includegraphics[height=5cm,keepaspectratio]{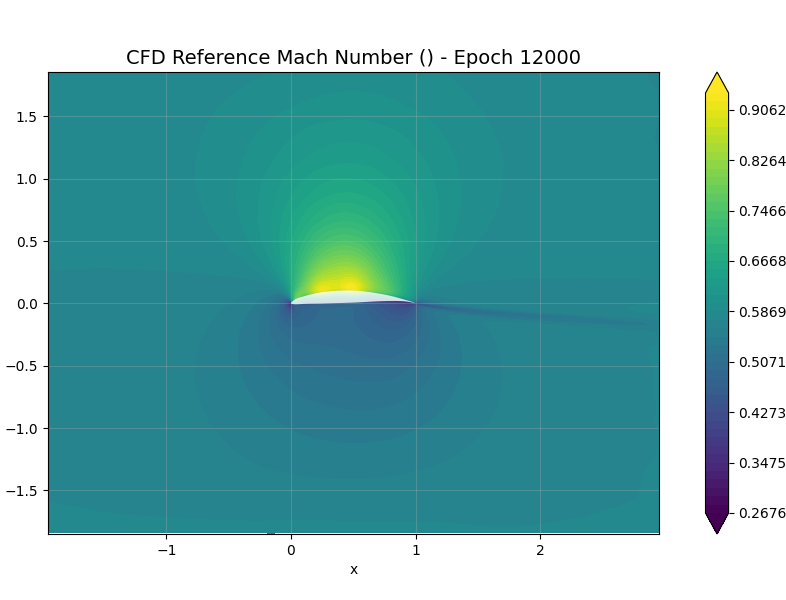}} &
      \subcaptionbox*{(i)\;\;$M$ Error}{%
        \includegraphics[height=5cm,keepaspectratio]{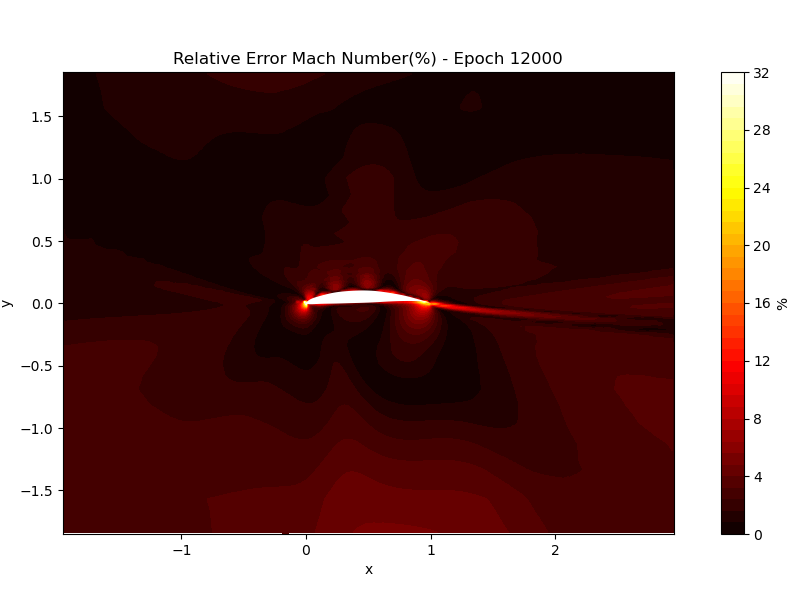}}
    \end{tabular}%
  }

  \caption{Airfoil 65-1010, \(\alpha = 5^{\circ}\), \(p_{\text{out}} = 0.80\,\text{bar}\).
           Rows: PINN prediction, SharC CFD, relative error.
           Columns: density \(\rho\), Mach number \(M\), static pressure \(p\).}
  \label{fig:cp_mach_1010A5}
\end{figure}

%----------------------------------------------------------------------

Figure \ref{fig:cp_curve_1010} compares the pressure coefficient distributions obtained from CFD and the proposed PINN framework. Overall, the PINN solution demonstrates a high level of accuracy on both the pressure side and the suction side, capturing the global pressure trends with good fidelity. 

On the pressure side, the agreement is nearly perfect over the upstream portion of the airfoil. However, discrepancies become more pronounced in the vicinity of the trailing edge. This deterioration can be primarily attributed to the geometric characteristics of the NACA65-1010 profile, which exhibits a downward curvature reversal near the trailing edge. Such rapidly varying geometric features introduce localised high-frequency components in the solution, which are known to be challenging for PINNs to resolve accurately. In particular, standard PINNs tend to struggle with sharp gradients and high-frequency behaviour due to spectral bias and optimisation stiffness, leading to reduced accuracy in these regions \cite{Krishnapriyan2021}. This effect is also reflected in the velocity field (see Fig \ref{fig:cp_mach_1010A5}), where deviations become more evident near the same region.

A secondary contributing factor arises from the geometric discontinuity at the trailing edge, where the network is required to simultaneously approximate the velocity field associated with both upper and lower surfaces. This effectively introduces competing constraints in a small spatial region, which can further degrade the solution quality. Similar behaviour is observed near the leading edge and in the suction-side region close to the trailing edge (Fig \ref{fig:cp_mach_1010A5}), where strong curvature variations are present.

On the suction side, the PINN provides an overall accurate prediction of the pressure distribution. However, it tends to smooth out fine-scale variations induced by local curvature changes. This behaviour is consistent with the well-documented tendency of neural networks, including PINNs, to favour smooth, low-frequency solutions and to under-represent high-frequency features or sharp transitions \cite{Krishnapriyan2021}. More generally, recent studies have shown that PINNs exhibit an intrinsic bias toward over-smoothed solutions and may struggle to capture sharp fronts or multi-scale structures in complex flows \cite{naser2025fundamental}.

Overall, despite the geometric complexity and the presence of curvature discontinuities in the NACA65-1010 profile, the PINN achieves a satisfactory level of accuracy, with a lift coefficient error of approximately $8.37\%$, demonstrating its robustness for moderately complex aerodynamic configurations.

%----------------------------------------------------------------------

%----------------------------------------------------------------------

\paragraph{Loss evolution}  
Table \ref{tab:time_breakdown} compares three strategies on the same cases(NACA1010 AoA $5\deg$, 0.8 bar outlet): (i) plain cold‑start PINN (diverges), (ii) geometry warm‑start only, and (iii) the proposed combined schedule.  The warm‑start run reduces the interior PDE residual but stalls once the slip‑wall and PDE losses conflict, while the combined approach continues
to converge, reaching \(R_{\max}<10^{-4}\) after only \(\sim4000\)
Adam iterations.

\paragraph{Stage‑wise runtime}  
Table\ref{tab:time_breakdown} lists Adam and L‑BFGS iterations together
with wall‑clock time on one NVIDIA A5000 GPU.

\begin{table}[H]
  \centering
  \caption{Convergence statistics—airfoil65‑0010,
           \(\alpha=0^{\circ}\), \(p_{\text{out}}=0.8\)bar.}
  \label{tab:time_breakdown}
  \begin{tabular}{@{}lrrrr@{}}
    \toprule
    Strategy & Adam iters & L‑BFGS iters & Time [min] & Outcome \\ \midrule
    Cold start        & 20\,000 & 0      & 35 & diverged \\
    Warm‑start only   & 10\,000 & 3\,000 & 18 & stalled  \\
    \textbf{Combined} & \textbf{4\,000} & \textbf{1\,800} & \textbf{12} & converged \\ \bottomrule
  \end{tabular}
\end{table}

\begin{table}[H]
  \centering
  \caption{Average speed-up relative to SharC CFD.}
  \label{tab:speed_up}
  \makebox[\textwidth][c]{%
    \begin{tabular}{@{}lccc@{}}
      \toprule
      Method & Cases & Hardware & Speed-up × \\ \midrule
      SharC CFD & NACA65-0310, AoA $0^\circ$, 0.8bar & CPU 6-core, $\sim$1.5h & 1 \\
      Cold-start PINNs & NACA65-0310, AoA $5^\circ$, 0.8bar & A5000 GPU, $\sim$43min & $\sim$2 \\
      \textbf{Progressive PINNs} & NACA65-0000 to 65-1010, AoA $0^\circ$, 0.8bar & A5000 GPU, $\sim$18min (avg) & \textbf{5} \\ \bottomrule
    \end{tabular}%
  }
\end{table}

%----------------------------------------------------------------------
% Horizontal triptych: Mach number fields
%----------------------------------------------------------------------

\paragraph{Industrial extrapolation}  
Projecting to 100 shapes (common in compressor design), the progressive
workflow would finish in $\approx$ 30 GPU-hours—two wall-days on a single A5000 
while CFD would require $\approx$ 150 CPU-hours.

%----------------------------------------------------------------------
\subsection{Design–matrix overview}\label{sec:res_matrix}
%----------------------------------------------------------------------

Figure \ref{fig:design_matrix_cl} shows a visual summary of all the test points used in this study plotted in terms of their aerodynamic design variables. On the vertical axis we have the design lift coefficient, which in the NACA-65 naming convention corresponds to the first digit after the series number (for example, a “3” in 65-0310 represents a design lift coefficient of 0.3). On the horizontal axis is the maximum thickness of the airfoil, expressed as a percentage of the chord length. Members of the same camber family, that is, airfoils with identical thickness, are connected by solid blue lines to help guide the eye through the design space. Each red circle in the figure represents a prediction made by the physics-informed neural network across the 20 combinations of geometry and operating conditions considered. In addition, the subset of cases that have been verified independently using our in-house SharC finite-volume solver is indicated with the special marker shown in black; these twelve cases provide reference data against which the PINN predictions can be compared. The size of each marker reflects the angle of incidence, with smaller markers corresponding to zero degrees and larger markers corresponding to five degrees. Finally, hollow markers are used to denote cases in which the outlet static pressure was set to 0.80 bar, distinguishing them from runs conducted at the baseline outlet condition.

% 需要的宏包：
% \usepackage{pgfplots}
% \pgfplotsset{compat=1.18}
% \usepackage{pgfplotstable}
% \usepackage{siunitx} % 可选，用于数值格式

%==================== DATA (replace CL_* numbers with your values) ====================
% columns: m  t  alpha  pout   src  cl
% alpha: 0 or 5 (deg)
% pout : 0.95 or 0.80 (bar)
% src  : 0 = PINN, 1 = CFD
% ================== 必要宏包（放在导言区） ==================
% \usepackage{tikz,pgfplots,pgfplotstable}
% \pgfplotsset{compat=1.18}

% ================== 颜色/样式定义 ==================
\tikzset{
  stylePINN095A/.style={red,   thick, mark=*,   mark size=1.8pt},
  stylePINN080A/.style={blue,  thick, dashed, mark=o, mark size=1.8pt},
  stylePINN095B/.style={green!60!black, thick, mark=*,   mark size=1.8pt},
  stylePINN080B/.style={violet,thick, dashed, mark=o, mark size=1.8pt},
  styleCFD/.style={
    only marks, mark=x, mark size=5 pt, line width=1.1pt,
    black, preaction={draw=white, line width=2.4pt}
}
}

% ================== 数据表：一行一个 m（= design lift m/10） ==================
% 说明：
% 1) 缺数据就填 "nan"，不要留空；这样 pgfplots 会跳过该点。
% 2) cl_pinn_* 列是 PINN；cl_cfd_* 列是 CFD 参照。
% 3) 下面只是示例数值，请替换成你的 5 组 airfoil 的结果。
\pgfplotstableread[col sep=space]{
m   cl_pinn_095_0  cl_pinn_080_0  cl_pinn_095_5  cl_pinn_080_5  cl_cfd_095_0  cl_cfd_080_0  cl_cfd_095_5  cl_cfd_080_5
0     -0.03           0.01           0.45           0.48           0.00          nan          nan           nan
3      0.34           0.28           0.62           0.65           nan           0.33          nan           0.67
6      0.58           0.54           0.89           0.95           0.61           nan          0.93          nan
10     nan            nan            1.03           1.14           nan           nan          nan           1.25
}\designdata

% ================== 图：按 (alpha, p_out) 连接 PINN，CFD 画“×”不连线 ==================

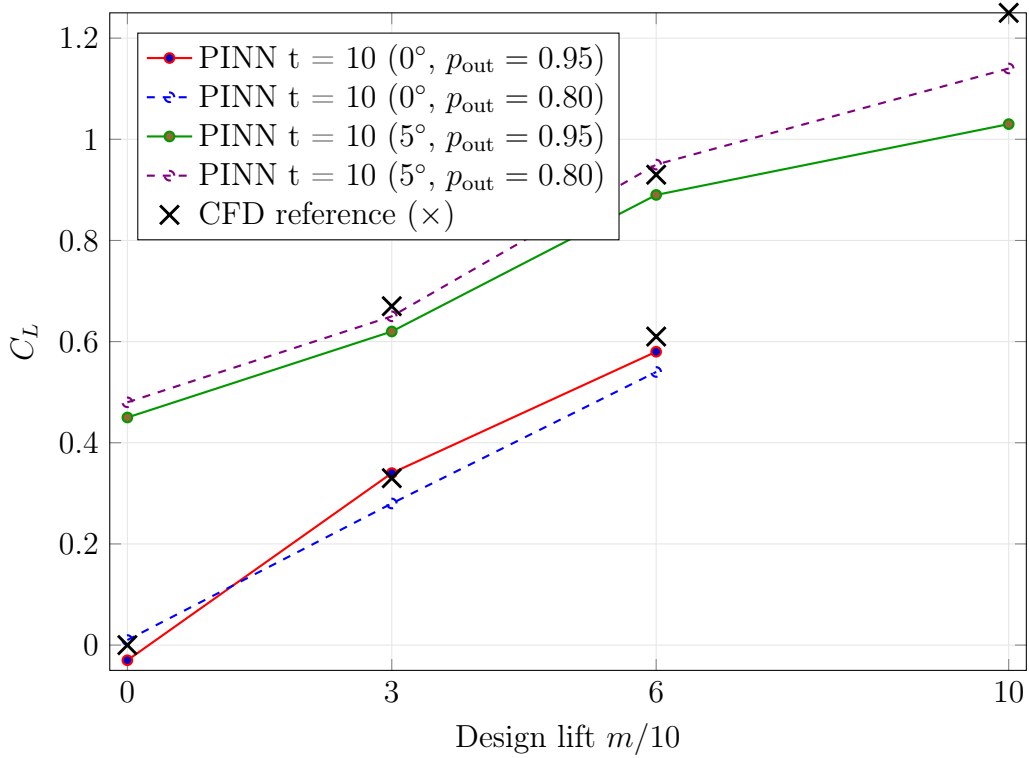
\begin{figure}[H]
  \centering
  \begin{tikzpicture}
    \begin{axis}[
      width=1\linewidth, height=0.75\linewidth,
      xmin=-0.2, xmax=10.2,  ymin=-0.05, ymax=1.25,
      xtick={0,3,6,10},
      xlabel={Design lift $m/10$},
      ylabel={$C_L$},
      legend pos=north west,
      legend cell align=left,
      grid=both, major grid style={gray!20},
      every axis plot/.append style={line join=round}
    ]

      % -------- PINN 四条曲线（连线） --------
      \addplot+[stylePINN095A]
        table[x=m, y=cl_pinn_095_0] {\designdata};
      \addlegendentry{PINN t = 10 ($0^\circ$, $p_{\text{out}}=0.95$)}

      \addplot+[stylePINN080A]
        table[x=m, y=cl_pinn_080_0] {\designdata};
      \addlegendentry{PINN t = 10 ($0^\circ$, $p_{\text{out}}=0.80$)}

      \addplot+[stylePINN095B]
        table[x=m, y=cl_pinn_095_5] {\designdata};
      \addlegendentry{PINN t = 10 ($5^\circ$, $p_{\text{out}}=0.95$)}

      \addplot+[stylePINN080B]
        table[x=m, y=cl_pinn_080_5] {\designdata};
      \addlegendentry{PINN t = 10 ($5^\circ$, $p_{\text{out}}=0.80$)}

      % -------- CFD 参照（只画散点 ×，不连线）--------
      \addplot+[styleCFD]
        table[x=m, y=cl_cfd_095_0] {\designdata};
      \addplot+[styleCFD]
        table[x=m, y=cl_cfd_080_0] {\designdata};
      \addplot+[styleCFD]
        table[x=m, y=cl_cfd_095_5] {\designdata};
      \addplot+[styleCFD]
        table[x=m, y=cl_cfd_080_5] {\designdata};
      \addlegendentry{CFD reference ($\times$)}

    \end{axis}
  \end{tikzpicture}
  \caption{$C_L$ vs.\ design lift $m/10$. PINN points are connected per $(\alpha,p_{\text{out}})$; CFD is shown as “$\times$” without lines. NACA1010 with AoA 0 doesn't have a steady state Euler solution}
  \label{fig:design_matrix_cl}
\end{figure}

\paragraph{Observations.}  
Our results highlight a clear hierarchy in how different design- and operating-variables influence the accuracy of the PINN predictions.  First and most significantly, blade geometry — in particular thickness and camber/curvature — dominates the variation in error: when the curvature and thickness remain moderate, the PINN framework maintains near-CFD fidelity, but its performance degrades appreciably as these increases.  Second, the incidence angle (e.g.\ comparing $0^\circ$ and $5^\circ$) has a measurable yet less dramatic effect on accuracy: higher angles of attack tend to increase errors, but the impact remains substantially smaller than that stemming from geometry changes.  Third, variations in the flow-field conditions (such as changes in freestream velocity or outlet static pressure, e.g.\ \(p_{\mathrm{out}} = 0.95\) vs.\ \(0.80\,\mathrm{bar}\)) produce only modest degradation, provided the geometry and incidence are fixed.  In other words, once geometry and incidence are fixed, the influence of operating-point variations is comparatively minor.  Taken together, these findings suggest that for rapid blade screening using our PINN workflow, the design domain characterised by moderate geometry (low curvature, moderate thickness) and moderate incidence represents a freindly combinaton of parameters where high accuracy is preserved.  For configurations involving highly curved geometry or high incidence, one must exercise increased caution — even modest changes in operating-point have limited effect compared to geometry-induced error drift.

%----------------------------------------------------------------------
%----------------------------------------------------------------------

\section{Conclusion}
This study has developed a progressive, geometry-aware PINN framework for two-dimensional aerodynamic screening of turbine-blade profiles — combining a boundary-progressive training schedule with a curvature-weighted residual loss to enable efficient, geometry-sensitive evaluation across an airfoil family. The results demonstrate that, within a broad set of 20 geometry and operating-point configurations, the method converges robustly and maintains agreement with CFD reference data within error margins on the order of 1\%.

Notably, blade geometry (camber and thickness/curvature) emerges as the dominant factor controlling prediction error, with incidence angle and operating-point variations (flow velocity or outlet pressure) exerting only secondary or tertiary influence. In regimes of moderate geometry and loading, our PINN workflow offers a favorable trade-off between speed and fidelity: evaluations complete on the order of tens of minutes on a single GPU, yet deliver accuracy approaching that of traditional finite-volume solvers. For strongly curved or heavily loaded configurations, persistent under-prediction of lift and degraded boundary-layer fidelity indicate that, at present, the method is best suited for preliminary screening — with final performance verification via high-fidelity CFD or an enhanced PINN protocol advised.

From an engineering-design perspective, this framework substantially lowers the computational barrier for early-phase blade family exploration, enabling rapid parametric sweeps over large design spaces. Looking ahead, improving robustness in high-curvature or high-loading cases (e.g.\ via advanced residual weighting, adaptive collocation, or multi-fidelity seeding), extending the approach to viscous (RANS/LES) flows, exploring systematic weight reuse across geometry families, and developing uncertainty quantification and error-certification mechanisms represent promising directions for future research.

Overall, our work shows that — when carefully tailored — PINNs can serve as viable, high-throughput surrogate models for blade aerodynamic pre-design. While not yet a full substitute for high-fidelity CFD in all regimes, they significantly expand the practical design-space that can be explored rapidly and inform design-decisions in early project phases.

In conclusion, the present study demonstrates that PINNs—when carefully tailored via geometry‑aware scheduling and loss‑design—can indeed serve as viable, high‑throughput surrogates for blade aerodynamic screening. While they cannot replace high‑fidelity CFD in all regimes, they significantly expand the accessible design‑space for rapid exploration and optimisation.

%% Refer following link for more details.
%% https://en.wikibooks.org/wiki/LaTeX/Mathematics
%% https://en.wikibooks.org/wiki/LaTeX/Advanced_Mathematics

%% Use a table environment to create tables.
%% Refer following link for more details.
%% https://en.wikibooks.org/wiki/LaTeX/Tables

%% Labels are used to cross-reference an item using \ref command.
%% The Appendices part is started with the command \appendix;
%% appendix sections are then done as normal sections

%% If you have bib database file and want bibtex to generate the
%% bibitems, please use
%%
%%  \bibliographystyle{elsarticle-num} 
%%  \bibliography{<your bibdatabase>}

%% else use the following coding to input the bibitems directly in the
%% TeX file.

%% Refer following link for more details about bibliography and citations.
%% https://en.wikibooks.org/wiki/LaTeX/Bibliography_Management
\bibliographystyle{plain} 
\bibliography{reference}

\end{document}